\documentclass{article}

\usepackage{spconf,amsmath,graphicx,hyperref}
\usepackage{booktabs}
\usepackage{multirow}
\usepackage{array}
\usepackage{bm}
\usepackage{xcolor}
\usepackage{amssymb}
\hypersetup{hidelinks}
\newcommand{\best}[1]{\textbf{#1}}
\newcommand{\second}[1]{\underline{#1}}

\title{QK-GCC: LEARNABLE QUERY-KEY SPECTRAL MATCHING FOR ROBUST TIME DELAY ESTIMATION}

\name{
Jinkai Zhang$^{1}$ \qquad
Weiye Chen$^{2}$ \qquad
Yue Huang$^{1,2}$ \qquad
Xiaotong Tu$^{1,2}$ \qquad
Xinghao Ding$^{1,2}$
}

\name{
	Jinkai Zhang$^{1}$ \qquad
	Weiye Chen$^{2}$ \qquad
	Yue Huang$^{1,2}$ \qquad
	Xiaotong Tu$^{1,2}$ \qquad
	Xinghao Ding$^{1,2}$
	\thanks{This work has been submitted to the IEEE for possible publication.
		Copyright may be transferred without notice, after which this version may
		no longer be accessible.}
}

\address{
$^{1}$School of Informatics, Xiamen University, Xiamen, China\\
$^{2}$Institute of Artificial Intelligence, Xiamen University, Xiamen, China\\
}

\begin{document}
\ninept
\maketitle

\begin{abstract}
Time delay estimation (TDE) is a fundamental component of microphone-array sound source localization. Generalized cross-correlation (GCC) is widely used because it is efficient and interpretable, but its handcrafted spectral matching and predefined frequency weighting are vulnerable to noise and reverberation. Existing neural GCC variants mainly improve robustness by enhancing input signals or modeling GCC responses, while the cross-channel spectral matching step itself remains handcrafted. We propose QK-GCC, a learnable GCC-like framework that replaces handcrafted weighted spectral matching in GCC with Query-Key matching between two microphone signals. The two microphone signals are encoded as magnitude-phase frequency tokens and mapped to Query and Key representations, respectively, enabling frequency reliability learning and local spectral evidence aggregation for delay estimation. Experiments in simulated reverberant rooms across diverse SNR and reverberation conditions show that QK-GCC improves TDE accuracy over GCC-PHAT and learning-based GCC variants, while remaining lightweight and generalizing to unseen source types. The code is available at \url{https://github.com/zhangjinkai33-ui/QK-GCC}.
\end{abstract}


\section{Introduction}

Time delay estimation (TDE), or time difference of arrival (TDOA) estimation, is a core component of microphone-array sound source localization. Since the source waveform and emission time are usually unknown, the relative delay between two microphones must be inferred from the similarity between their received signals. Accurate delay estimates provide essential spatial cues for downstream direction-of-arrival estimation and source tracking~\cite{knapp1976generalized,diazguerra2020tracking,li2016reverberant,burgess2015toa,astrom2021extension}.

Generalized cross-correlation (GCC) is among the most widely used TDE frameworks due to its simple formulation, efficient implementation, and clear physical meaning~\cite{knapp1976generalized}. At its core, GCC estimates delay by accumulating cross-channel spectral evidence across frequencies under each delay hypothesis. GCC-PHAT, its dominant variant, normalizes the cross-spectrum so that the delay response mainly depends on inter-channel phase differences. Other weighting schemes, such as Roth, SCOT, and maximum-likelihood weighting, encode different assumptions about the signal and noise spectra~\cite{roth1971effective,carter1973smoothed,hannan1973estimating,donohue2007performance}. However, these weightings are predefined and cannot adapt to the reliability of each frequency bin in the current input. Under noise and reverberation, corrupted spectral components may therefore broaden, shift, or split the correlation peak, leading to inaccurate delay estimates.

Recent learning-based methods improve TDE robustness by
introducing neural processing into the GCC pipeline, including
time-frequency masking, sub-band GCC extraction, parametrized
GCC-PHAT responses, multi-GCC fusion, neural filtering before
GCC-PHAT, and learned filtering of GCC responses~\cite{wang2018robust,wang2019robust,wang2021gcc,cobos2020frequency,comanducci2020time,salvati2021time,liu2023time,berg2022extending,berg2024multitarget,gulin2024gcc}. Recent probing further shows that neural TDOA estimators learn
magnitude-aware frequency weighting rather than PHAT whitening
~\cite{kang2026what}. In GCC-based pipelines, however, learning is mostly applied before or
after the GCC operation, while the core cross-channel spectral matching
used to form the GCC response remains handcrafted. This reveals a fundamental limitation of handcrafted GCC-style matching: the reliability weighting of frequency bins and the aggregation of neighboring spectral evidence are both fixed by design. As a result, the matching rule cannot learn data-driven frequency reliability, and its strict bin-to-bin form cannot exploit nearby evidence when individual bins are corrupted. These limitations make the delay response less reliable in adverse acoustic
conditions, especially under strong noise and reverberation.

To address these limitations, we propose QK-GCC, which replaces fixed weighted
spectral matching in GCC-style TDE with a local Query-Key matching operator. The two signals are represented by magnitude-phase frequency tokens and mapped to Query and Key representations, respectively, before being matched within local frequency neighborhoods to produce delay-discriminative responses. QK-GCC therefore keeps the GCC-style view of estimating delay from cross-channel spectral matching, while replacing the fixed matching rule with a learned operator. By learning how two spectra should be matched, rather than only processing signals or GCC responses, QK-GCC targets the core limitation of handcrafted GCC-style correlation while remaining a lightweight two-channel operator.

The main contributions of this paper are summarized as follows:

\begin{itemize}
\item We propose QK-GCC, a lightweight GCC-like TDE operator that replaces handcrafted weighted spectral matching in GCC with learnable local Query-Key matching between two microphone spectra.

\item We decompose the Query-Key similarity into a learned
frequency-reliability factor and an angular similarity term,
providing learnable counterparts to GCC frequency weighting
and cross-channel matching.

\item We demonstrate that QK-GCC improves delay estimation under noise, reverberation, and unseen source types while remaining lightweight, and that its learned reliability score aligns with the average inter-channel coherence profile.
\end{itemize}

\section{Method}

\begin{figure*}[t]
\centering
\includegraphics[width=0.98\textwidth]{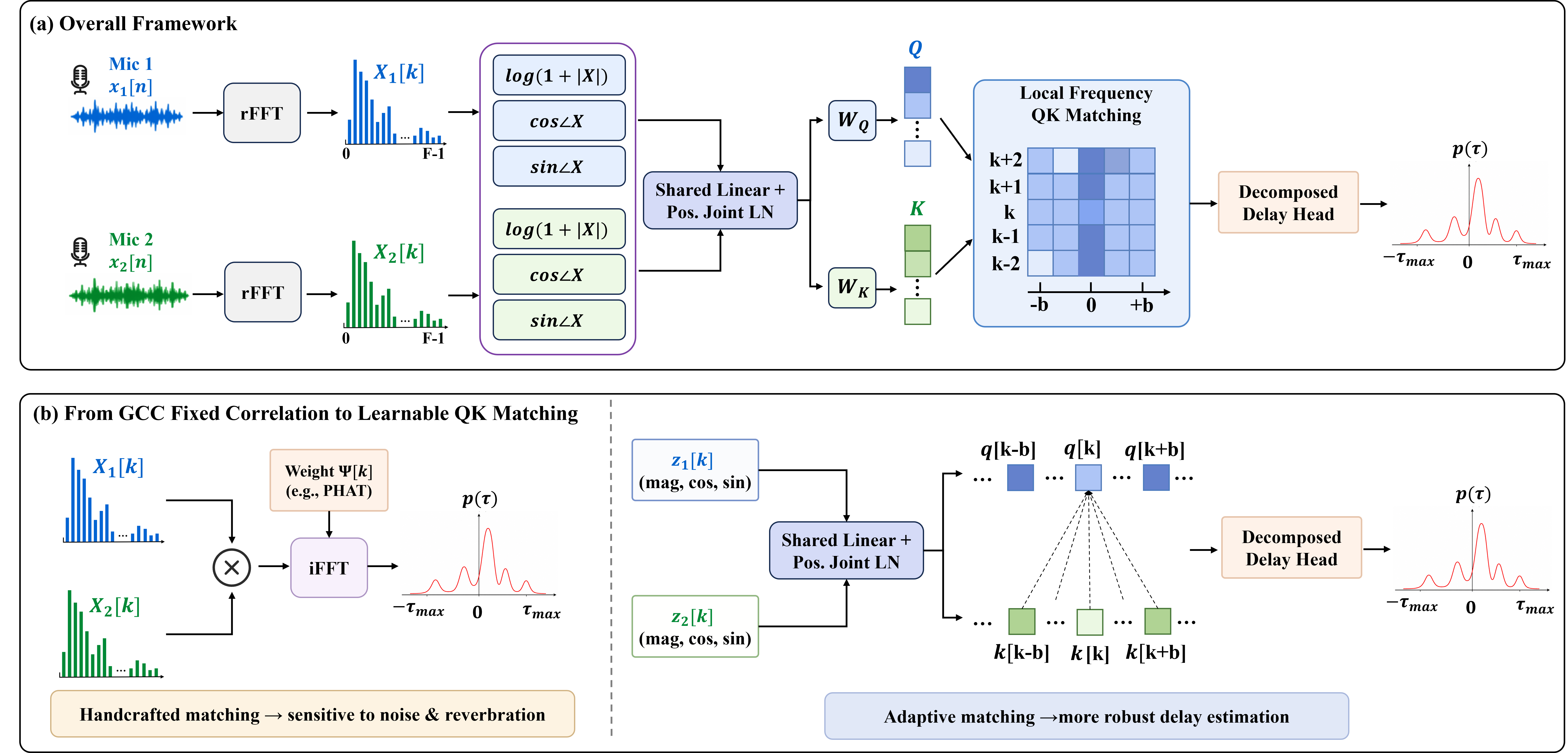}
\caption{Overall architecture of QK-GCC. Conventional GCC forms a delay response from a handcrafted cross-channel spectral product with predefined frequency weighting, whereas QK-GCC replaces this fixed spectral matching with learnable local Query-Key matching in the frequency domain.}
\label{fig:structure}
\end{figure*}

\subsection{Revisiting GCC as Weighted Spectral Matching}

Given two synchronized microphone signals \(x_1[n]\) and \(x_2[n]\), time delay estimation aims to infer their relative delay \(\tau\). Generalized cross-correlation estimates the delay by searching for the peak of
a correlation function formed from a weighted cross-spectrum:
\begin{equation}
R_{12}[\tau]
=
\sum_{k=0}^{K-1}
\Psi[k] X_1[k] X_2^{*}[k]
e^{j2\pi k\tau/K},
\label{eq:gcc}
\end{equation}
where \(X_1[k]\) and \(X_2[k]\) are the spectra of the two signals,
\(K\) is the DFT length, and \(\Psi[k]\) is a predefined frequency weighting.
The term \(\Psi[k]X_1[k]X_2^*[k]\) provides frequency-wise spectral evidence,
which is accumulated across frequencies after modulation by each delay
hypothesis.

Writing \(X_m[k]=|X_m[k]|e^{j\phi_m[k]}\), this spectral evidence decomposes into a spectral weighting and a phase matching term:
\begin{equation}
\Psi[k]X_1[k]X_2^*[k]
=
\underbrace{\Psi[k]|X_1[k]||X_2[k]|}_{\text{spectral weighting}}
\underbrace{e^{j(\phi_1[k]-\phi_2[k])}}_{\text{phase matching}} .
\label{eq:gcc_decomp}
\end{equation}
Thus, GCC-style TDE relies on a handcrafted frequency weighting and a fixed cross-channel phase-matching rule. QK-GCC replaces both with a learnable Query-Key similarity, as summarized in Fig.~\ref{fig:structure} and described next.

\subsection{QK-GCC: Learnable Spectral Matching}

\noindent\textbf{Learnable matching.}~
We replace the handcrafted frequency-wise matching rule in
Eq.~\eqref{eq:gcc} with learnable Query-Key matching:
\begin{equation}
\Psi[k]X_1[k]X_2^{*}[k]
\quad\Longrightarrow\quad
\langle \mathbf{q}_1[k], \mathbf{k}_2[k]\rangle ,
\label{eq:qk_replace}
\end{equation}
where \(\mathbf{q}_1[k]\) and \(\mathbf{k}_2[k]\) are Query and Key representations derived from the two signals, constructed as described in the following paragraph. The resulting similarities serve as learnable frequency-wise evidence and are later aggregated by the delay head. The inner product decomposes as
\begin{equation}
\langle \mathbf{q}_1[k], \mathbf{k}_2[k]\rangle
=
\|\mathbf{q}_1[k]\|\,\|\mathbf{k}_2[k]\|\cos\theta_k .
\label{eq:qk_decomp}
\end{equation}
Here, the norm product \(\|\mathbf{q}_1[k]\|\,\|\mathbf{k}_2[k]\|\) can be interpreted as playing the weighting role of \(\Psi[k]|X_1[k]||X_2[k]|\) in Eq.~\eqref{eq:gcc_decomp}, while \(\cos\theta_k\) plays a role analogous to the cross-channel matching term. The Query-Key inner product therefore provides a learnable analogue of the two-factor structure in GCC spectral evidence, with representation magnitudes reflecting reliability-like weighting and angular similarity reflecting cross-channel matching. This interpretation is enabled by the GCC-inspired construction of frequency-domain Query-Key matching. Unlike standard attention~\cite{vaswani2017attention}, QK-GCC uses the
Query-Key score directly as cross-channel delay evidence, with no
Value projection.

\noindent\textbf{Query and Key construction.}~
The spectrum of each signal is computed as \(X_m[k]=\mathrm{rFFT}(x_m[n])\). We denote by \(F\) the number of retained non-negative frequency bins after rFFT. Each bin is represented by magnitude and phase features:
\begin{equation}
\mathbf{z}_m[k]
=
\left[
\log(1+|X_m[k]|),
\cos\angle X_m[k],
\sin\angle X_m[k]
\right].
\label{eq:token}
\end{equation}
The log magnitude serves as a reliability cue, while the cosine-sine phase representation avoids phase wrapping.

The two signals are mapped into a shared embedding space and then projected into Query and Key representations:
\begin{equation}
\mathbf{h}_m[k]
=
\mathrm{LN}\left(\mathbf{z}_m[k]\mathbf{W}_{in}+\mathbf{e}_k\right),
\label{eq:frontend}
\end{equation}
\begin{equation}
\mathbf{q}_1[k]=\mathbf{h}_1[k]\mathbf{W}_Q,
\qquad
\mathbf{k}_2[k]=\mathbf{h}_2[k]\mathbf{W}_K ,
\label{eq:qk}
\end{equation}
where \(\mathbf{W}_{in}\) is shared across the two channels, \(\mathbf{e}_k\) is a learnable frequency positional embedding, and \(\mathrm{LN}\) denotes layer normalization.

\subsection{Local Frequency Correlation and Delay Head}
\label{ssec:local}
Equation~\eqref{eq:qk_replace} defines learnable evidence for aligned frequency bins. To relax the strict bin-to-bin matching used in conventional GCC, QK-GCC further extends the matching to a local frequency neighborhood of radius \(b\), allowing evidence to be collected from neighboring bins under noise, reverberation, and spectral leakage. This local correlation is computed across \(H\) parallel heads. For the \(h\)-th head, it is defined as
\begin{equation}
C_h[k,\delta]
=
\frac{\exp(\alpha_h)}{\sqrt{d_h}}
\left\langle
\mathbf{q}_{1,h}[k],
\mathbf{k}_{2,h}[k+\delta]
\right\rangle,
\quad
\delta\in[-b,b],
\label{eq:local_qk}
\end{equation}
where \(d_h\) is the dimension of each head and \(\exp(\alpha_h)\) is a learnable positive scale. Boundary frequency bins are handled by zero padding. This produces a local correlation tensor \(\mathbf{C}\in\mathbb{R}^{H\times(2b+1)\times F}\).
To turn the local correlation into delay logits, we decompose the offsets into center, symmetric, and antisymmetric components:
\begin{equation}
\begin{aligned}
C_h^{0}[k] &= C_h[k,0],\\
C_h^{\mathrm{sym}}[k,\delta]
&= \frac{1}{2}\left(C_h[k,\delta]+C_h[k,-\delta]\right),\\
C_h^{\mathrm{asym}}[k,\delta]
&= \frac{1}{2}\left(C_h[k,\delta]-C_h[k,-\delta]\right),
\end{aligned}
\label{eq:sym}
\end{equation}
for \(\delta>0\). The center term provides bin-aligned evidence; the symmetric term captures local consistency shared across neighboring frequencies; the antisymmetric term captures directional asymmetry across offsets. The three components are concatenated along the offset axis and fed to a
lightweight delay head comprising two 1-D convolutional blocks, frequency
pooling, and an MLP that outputs logits over the candidate delay grid. Unless otherwise specified, QK-GCC uses an embedding dimension of
\(d=128\), \(H=4\) heads (\(d_h=32\)), and a local bandwidth of \(b=8\).

\subsection{Training Objective}

Following~\cite{berg2022extending}, we formulate frame-level TDE as classification over the delay grid
\(\mathcal{T}=\{-\tau_{\max},\ldots,\tau_{\max}\}\). The model is trained with cross-entropy, and inference selects the delay with the maximum posterior probability.

\section{Experiments}

\subsection{Experimental Setup}

We simulate two-channel signals with Pyroomacoustics~\cite{scheibler2018pyroomacoustics}
using the image-source method. Clean speech is drawn from
LibriSpeech~\cite{panayotov2015librispeech} (\(16\,\mathrm{kHz}\)),
with speaker-disjoint training, validation, and test sets. After voice-activity-based silence removal, we extract 2048-sample frames
with a 4096-point FFT and formulate TDE as classification over
\([-\tau_{\max},\tau_{\max}]\), with \(\tau_{\max}=23\).
Training and test rooms are \(7{\times}5{\times}3\,\mathrm{m}\) and
\(6{\times}4{\times}2.5\,\mathrm{m}\), respectively, with different
array locations and \(0.5\,\mathrm{m}\) microphone spacing. Source
positions are uniformly sampled within each room. During training,
\(T_{60}\sim\mathcal{U}(0.2,1.0)\,\mathrm{s}\) and
SNR\(\sim\mathcal{U}(0,30)\,\mathrm{dB}\), with additive white noise,
yielding about \(1.1\times10^{5}\) training pairs. Random channel
swapping with delay-sign reversal is also used. Testing covers
SNRs \(\{0,6,12,18,24,30\}\,\mathrm{dB}\) and
\(T_{60}\in\{0.2,0.4,0.6,0.8,1.0\}\,\mathrm{s}\).
For zero-shot evaluation, source signals are replaced by
MUSAN~\cite{snyder2015musan} music and noise clips without changing
the acoustic simulation. All learnable models are trained for 30 epochs
using Adam with batch size 32, initial learning rate \(10^{-3}\), and
cosine decay.

We compare against GCC-PHAT~\cite{knapp1976generalized}, PGCC-PHAT~\cite{salvati2021time}, and NGCC-PHAT~\cite{berg2022extending}, spanning fixed correlation, parametrized GCC, and neural-filtering GCC. For a fair comparison, all learning-based methods are retrained from scratch on the same training pairs with the same optimizer, schedule, epochs, delay grid, and evaluation protocol, with no test-room or interference-specific fine-tuning. Since TDE is formulated as delay-grid classification here, PGCC-PHAT
is evaluated with its classification head, as adopted for NGCC-PHAT;
method-specific architectures otherwise follow their original papers.
FLOPs are estimated per 2048-sample input pair, with one
multiply--accumulate (MAC) counted as two FLOPs.
For compactness, GCC-PHAT, PGCC-PHAT, and NGCC-PHAT are abbreviated
as GCC, PGCC, and NGCC, respectively, in figures and tables.

\subsection{Evaluation Metrics}

We report RMSE and MAE in samples between the estimated delay
$\hat{\tau}$ and the ground truth $\tau$, together with Acc@10cm,
the percentage of predictions whose propagation-distance error
$e_d=|\hat{\tau}-\tau|c/f_s$ falls within 10 cm, where
$c=343\,\mathrm{m/s}$ is the speed of sound. Because downstream localization relies on most frames being near-correct, we treat MAE and Acc@10cm as the primary indicators of localization quality, and report RMSE to reflect sensitivity to large deviations.

\subsection{Main Results}

Table~\ref{tab:main_results} shows that QK-GCC obtains the lowest errors and the highest Acc@10cm while using the fewest parameters among learning-based methods. The improvements in both MAE and Acc@10cm indicate more consistent frame-level accuracy. The gain over NGCC-PHAT comes with markedly lower cost: roughly a third of the parameters and an order of magnitude fewer FLOPs, suggesting that directly learning the matching adds value beyond neural filtering ahead of a fixed GCC response.

\begin{table}[!t]
\centering
\caption{Overall TDE performance over three seeds, reported as mean (std).
RMSE and MAE are in samples.}
\label{tab:main_results}

\setlength{\tabcolsep}{2.0pt}
\renewcommand{\arraystretch}{1.02}

\begin{tabular}{@{}lccccc@{}}
\toprule
Method & Params & FLOPs & RMSE $\downarrow$ & MAE $\downarrow$ & Acc@10cm (\%) $\uparrow$ \\
\midrule

GCC
& --
& \best{0.7M}
& 11.89
& 8.00
& 48.57 \\

PGCC
& 11.56M
& 219M
& 10.44 (0.36)
& 6.37 (0.27)
& 57.85 (1.76) \\

NGCC
& \second{888K}
& 3682M
& \second{10.34 (0.08)}
& \second{6.06 (0.08)}
& \second{61.62 (0.47)} \\

QK-GCC
& \best{322K}
& \second{208M}
& \best{8.83 (0.07)}
& \best{5.29 (0.05)}
& \best{66.69 (0.27)} \\

\bottomrule
\end{tabular}
\end{table}

\subsection{Robustness under Noise and Reverberation}

\begin{figure}[!tbp]
\centering
\includegraphics[width=\columnwidth]{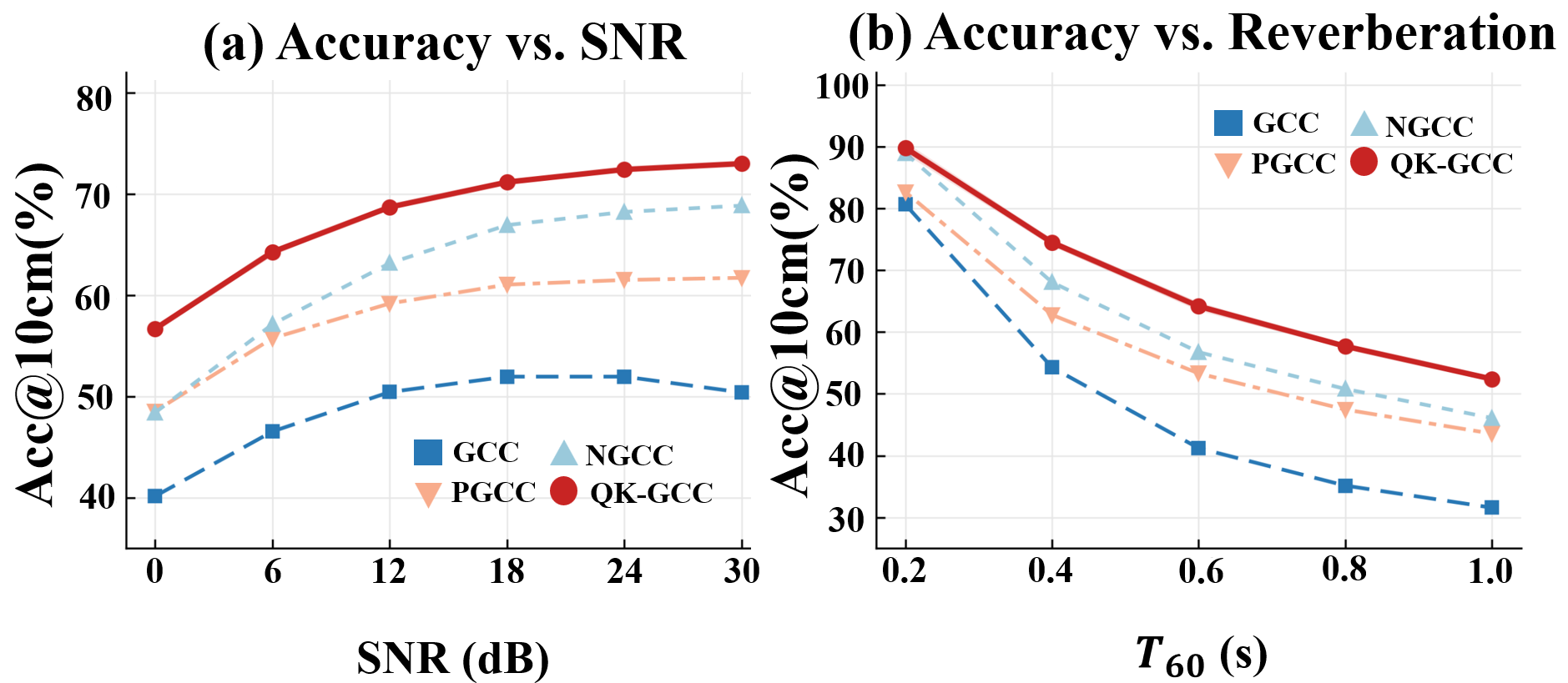}
\caption{Acc@10cm under varying (a) SNR averaged over all \(T_{60}\) and (b) reverberation time averaged over all SNR.}
\label{fig:acc_curves}
\end{figure}

Fig.~\ref{fig:acc_curves} shows that the advantage of QK-GCC becomes
more pronounced under harder acoustic conditions. Its margin widens as
SNR decreases, while its performance degrades more slowly as reverberation
increases. These trends agree with the roles of learned frequency-reliability
weighting and local frequency aggregation in QK-GCC.

\subsection{Zero-shot Generalization to Unseen Source Types}

We further test on MUSAN music and noise using the original checkpoints without fine-tuning.

\begin{table}[!t]
\centering
\caption{Zero-shot TDE with unseen MUSAN noise and music clips as source
signals (same acoustic simulation), over three seeds, reported as mean (std).
RMSE and MAE are in samples.}
\label{tab:zero_shot_musan}

\renewcommand{\arraystretch}{1.06}

\begin{tabular*}{\columnwidth}
{@{\extracolsep{\fill}}clccc@{}}
\toprule
& Method
& RMSE $\downarrow$
& MAE $\downarrow$
& Acc@10cm (\%) $\uparrow$ \\
\midrule

\multirow{4}{*}{\rotatebox[origin=c]{90}{\textit{Noise}}}
& GCC
& 11.75
& 8.01
& 47.48 \\

& PGCC
& 11.12 (0.39)
& 6.95 (0.25)
& 55.34 (1.38) \\

& NGCC
& \second{10.49 (0.08)}
& \second{6.35 (0.08)}
& \second{59.74 (0.28)} \\

& QK-GCC
& \best{9.37 (0.04)}
& \best{5.59 (0.02)}
& \best{65.46 (0.15)} \\

\cmidrule(lr){2-5}

\multirow{4}{*}{\rotatebox[origin=c]{90}{\textit{Music}}}
& GCC
& 12.46
& 8.70
& 44.52 \\

& PGCC
& 11.49 (0.21)
& 7.49 (0.19)
& 51.13 (1.46) \\

& NGCC
& \second{11.13 (0.07)}
& \second{7.13 (0.06)}
& \second{54.42 (0.36)} \\

& QK-GCC
& \best{9.59 (0.09)}
& \best{6.04 (0.03)}
& \best{60.54 (0.08)} \\

\bottomrule
\end{tabular*}

\end{table}

Table~\ref{tab:zero_shot_musan} shows that the trend in the main results also holds for source types never seen during training: QK-GCC keeps its lead on both MUSAN music and noise without adaptation. This is consistent with the model learning reusable cross-channel spectral correspondence rather than relying only on speech-specific patterns.

\subsection{Interpretability Analysis}
\label{sec:interp}

\begin{figure}[!t]
\centering
\includegraphics[width=\columnwidth]{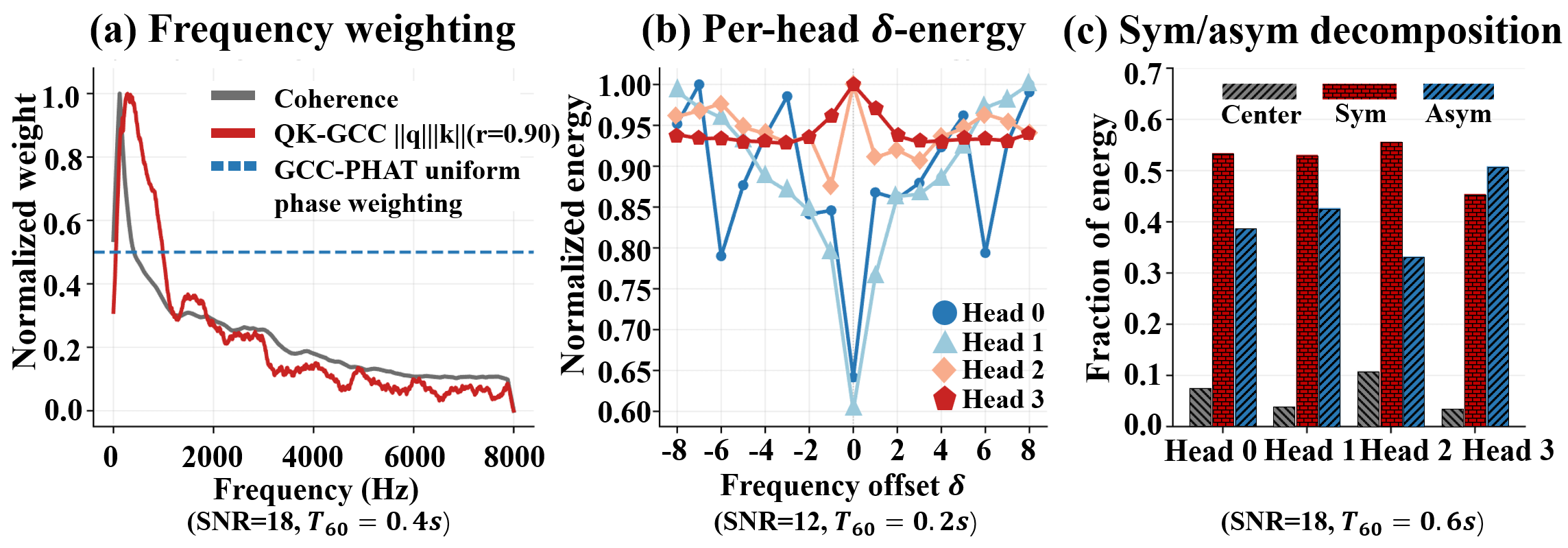}
\caption{Interpretability analysis of QK-GCC.
(a) Learned norm weighting and inter-channel coherence;
(b) normalized per-head energy across frequency offsets;
(c) center, symmetric, and antisymmetric component shares.}
\label{fig:interpretability}
\end{figure}

To examine what QK-GCC learns, we use the decomposition
$\langle \mathbf{q}_1[k], \mathbf{k}_2[k]\rangle
= \|\mathbf{q}_1[k]\|\,\|\mathbf{k}_2[k]\|\cos\theta_k$.
The norm product can be interpreted as a frequency-reliability score,
while the angular term reflects cross-channel matching.

For Fig.~\ref{fig:interpretability}(b), the per-head energy at each
offset is computed from the squared local-correlation response
$C_h[k,\delta]^2$, averaged over samples and frequency bins, and
normalized by the maximum across offsets for each head.
For Fig.~\ref{fig:interpretability}(c), we apply the decomposition in Eq.~\eqref{eq:sym} to each sample,
compute the squared energy of each component, and average it over samples.
The normalized energy shares are then reported for each head.

Fig.~\ref{fig:interpretability}(a) shows that, although trained without
coherence supervision, the learned norm weight
$\|\mathbf{q}_1\|\,\|\mathbf{k}_2\|$ closely follows the average
inter-channel coherence profile (Pearson $r=0.90$).
Fig.~\ref{fig:interpretability}(b) further shows that local QK matching
uses neighboring frequency offsets rather than only the zero offset,
while Fig.~\ref{fig:interpretability}(c) shows non-negligible
contributions from both symmetric and antisymmetric components.
Together, these results indicate that the GCC-structured operator guides
learning toward physically meaningful spectral matching.

\subsection{Ablation Study}

\begin{table}[!t]
\centering
\caption{Ablation study of QK-GCC (one representative seed).
RMSE and MAE are in samples.}
\label{tab:ablation}

\setlength{\tabcolsep}{3pt}
\renewcommand{\arraystretch}{1.02}

\begin{tabular}{@{}lccc@{}}
\toprule
Method & RMSE $\downarrow$ & MAE $\downarrow$ & Acc@10cm (\%) $\uparrow$ \\
\midrule
GCC                      & 11.89 & 8.00 & 48.57 \\
PGCC                     & 10.32 & 6.35 & 57.96 \\
NGCC                     & 10.25 & 5.98 & 62.12 \\
\midrule
Generic Transformer      & 10.12 & 6.76 & 52.35 \\
\midrule
QK-GCC w/o sym./asym.    & 9.08  & 5.60 & 64.08 \\
QK-GCC w/o magnitude     & 9.29  & 5.73 & 62.48 \\
QK-GCC w/ b=0         & 9.58  & 5.96 & 60.99 \\
QK-GCC w/ b=8         & \best{8.76}  & \best{5.26}  & \best{66.90} \\
QK-GCC w/ b=16        & \second{8.80} & \second{5.32} & \second{66.20} \\
QK-GCC w/ b=32        & 9.13  & 5.55 & 63.85 \\
\bottomrule
\end{tabular}
\end{table}

Table~\ref{tab:ablation} reports results from one representative seed.
A generic spectral Transformer with more parameters (418K vs.\ 322K),
using the same magnitude-phase tokens and delay grid but replacing the
cross-channel QK operator with standard self-attention, reaches only
52.35 Acc@10cm. This shows that the gain is not explained by model capacity alone and
highlights the importance of the GCC-inspired cross-channel matching
structure.
Removing magnitude cues reduces Acc@10cm from 66.90 to 62.48,
showing that phase-only matching is insufficient and that magnitude
provides useful reliability information. Removing the symmetric and
antisymmetric components lowers accuracy to 64.08, indicating that the explicit offset decomposition serves as a useful
structural prior for the delay head.
For the local bandwidth, $b=0$ drops to 60.99, while $b=16$ remains close
to $b=8$ (66.20 vs.\ 66.90) and $b=32$ falls to 63.85, showing that
local aggregation is most effective within a moderate frequency neighborhood.

\section{Conclusion}

We presented QK-GCC, a learnable GCC-like operator that replaces handcrafted weighted spectral matching with local Query-Key similarity. Its inner product decomposes into a learned frequency-reliability
score and an angular similarity term, providing learnable
counterparts to the weighting and cross-channel matching components
of GCC spectral evidence. QK-GCC outperforms GCC-PHAT and learning-based GCC variants under noise, reverberation, and unseen source types, while remaining lightweight and interpretable. Validation on real recordings and extension to multi-microphone arrays are left as future work.

\bibliographystyle{IEEEbib}
\bibliography{refs}

\end{document}